# Unexpected Collisional Rotational Excitation via Long-Range Capture and Orbiting

Dasarath Swaraj[1,#], Guodong Zhang[2, 3,#], Siting Hou[4,#], Dandan Lu[5,#], Jerin Judy[1], Fabio Zappa[1], Tim Michaelsen[1], Arnab Khan[1,+], Hua Guo[6,*], Changjian Xie[4,*], Hong Gao[2, 3,*], and Roland Wester[1,*]

[1]*Institut für Ionenphysik und Angewandte Physik, Universität Innsbruck, Technikerstraße 25, 6020 Innsbruck, Austria*

[2]*Beijing National Laboratory for Molecular Sciences (BNLMS), Institute of Chemistry, Chinese Academy of Sciences, Beijing 100190, China*

[3]*University of Chinese Academy of Sciences, Beijing 100049, China*

[4]*Institute of Modern Physics, Shaanxi Key Laboratory for Theoretical Physics Frontiers, Northwest University, Xi'an, Shaanxi 710127, China*

[5]*State Key Laboratory of High Temperature Gas Dynamics, Institute of Mechanics, Chinese Academy of Sciences, 100190 Beijing, China*

[6]*Department of Chemistry and Chemical Biology, Center for Computational Chemistry, University of New Mexico, Albuquerque, New Mexico 87131, USA*

[#] These authors contributed equally to this work.

[+] Present Address: Department of Physics, Indian Institute of Science Education and Research (IISER) Bhopal, Indore By-pass Road, Bhopal - 462066, India

* Corresponding authors. Email: hguo@unm.edu, chjxie@nwu.edu.cn; honggao2017@iccas.ac.cn, roland.wester@uibk.ac.at

**Abstract**

Collisional rotational excitation is a fundamental process in many gaseous environments. The textbook hard-sphere model stipulates that high rotational excitation results from head-on collisions, leading primarily to backward scattering, whereas long-range glancing collisions in the forward direction are inefficient for rotational energy transfer. Here, we report rotational state resolved product imaging for a system with strong attractive interaction, the charge-transfer collision between spin-orbit selected $Ar^+(^2P_{3/2})$ ions and *para/ortho*-$H_2$ molecules. Surprisingly, the $H_2^+$ products are rotationally excited and dominated by forward scattering, in sharp contrast to conventional wisdom. Quantum dynamical calculations on a first-principles diabatic potential energy matrix reproduce the observations. Trajectory surface hopping analysis further reveals that rotational excitation occurs mostly with large impact parameters, and the captured complex undergoes orbiting motion owing to the strong attractive interaction between the two collision partners before they break up. This novel mechanism should be general for collisional systems featuring strong attractive interactions, which undermine the hard-sphere assumption.

Collisional rotational excitation is a fundamental process in many gaseous environments, such as interstellar media, planetary atmospheres, and combustion. Numerous studies have measured rotationally inelastic collision cross sections (*1-3*), some at a quantum state-to-state level (*4-6*). Such information is particularly relevant in the absence of local thermodynamic equilibrium (*7*). Rotational excitation typically results from head-on collisions with near-zero impact parameters, much like those between billiard balls resulting in backward scattering, while large-impact-parameter glancing collisions in the forward direction are inefficient in inducing rotational excitation. Such a hard-sphere model, augmented with weak ubiquitous long-range attraction, is the established paradigm for understanding molecular collision dynamics (*8*). In this limit, collisional rotational excitation is essentially attributed to the anisotropic short-range repulsion, which exerts torque forces on the molecule, leading to rotational excitation. On the other hand, the long-range weak attractive forces only affect the collisions with impact parameters larger than the hard-sphere radius, and result in the so-called glory and L-type rainbow effects (*8*).

Recently, van de Meerakker and co-workers identified two new rotational excitation mechanisms in neutral systems that depart significantly from the hard-sphere archetype. These two mechanisms can be attributed to the existence of relatively strong attractive intermolecular interactions. In the hard-collision glory scattering (HCGS) mechanism, significant rotational excitation occurs through low-impact-parameter hard collisions, and the slowly recoiling products, which otherwise are expected to scatter sideways, are redirected to the forward direction because the long-range attraction is capable of balancing the repulsion, similar to the glory effect (*6*). The soft-collision backward glory scattering (SCBGS) mechanism, on the other hand, depicts backward scattering in nearly elastic long-range soft collisions with low rotational excitation, which are supposed to be forward scattered in the traditional model. Such backward scattering is found to follow a U-turn trajectory, owing to the relatively strong attractive interactions between the collision partners (*9*). Even though these phenomena deviate from the textbook hard-sphere model, they still conform to the basic belief that high rotational excitation results from hard collisions dominated by the repulsive part of the interaction potential (*8*).

These newly discovered mechanisms underscore the potential importance of attractive forces, which begs the question: how does collisional rotational excitation happen in systems that are dominated by attraction? So far, this question has rarely been addressed due to challenges in both experiments and theory. On the experimental side, most previously studied collisional systems feature relatively weak long-range attractive interactions. In this case an extremely low collision energy, which is experimentally challenging to create (*10, 11*), is needed for the attractive forces to exert a significant effect on a collision. On the other hand, strong long-range attractive interaction makes full-dimensional quantum scattering calculations difficult, especially for molecule-molecule collisions (*12*). Costen and coworkers recently performed a crossed-beam velocity-map imaging study on the collision between $CO_2$ and $NO(A^2\Sigma^+)$ (*13*) and observed that the rotationally excited products are dominantly scattered in the forward direction with a minor peak in the backward direction. The possibility of HCGS was firmly ruled out, because the collision energy is comparable to the depth of the attractive well in the interaction potential. Unfortunately, the full-dimensional potential energy surface and quantum scattering calculations on this system are not yet available, hindering our understanding of the rotational excitation mechanism in this case. In another example, Xie and coworkers recently carried out full-dimensional quantum scattering calculations on collisions between two HF molecules, which feature strong attractive interactions (*14*). At collision energies smaller than the depth of the potential well, they found that the scattered HFs are strongly rotationally excited, violating the well-known energy gap law, which is based on the hard-sphere model (*15*). This interesting prediction has not been verified by experiment yet, and the detailed mechanism remains elusive.

Collisions between ions and neutral molecules feature strong attractive interactions, thus are ideal for exploring novel mechanisms for collisional rotational excitation. While charge transfer is facile, it breaks no bond. As a result, such processes also provide a prototype for understanding rotational inelasticity. Until very recently, difficulties in preparing state-selected reactants and the relatively poor energy resolution in product detection for low-energy ion-molecule collisions have posed major obstacles. Of late, the adoption of quantum state-selected ion sources using laser photoionization and the advancement of ion-molecule crossed-beam setup based on three-dimensional velocity-map imaging have greatly mitigated these problems and proven powerful for unraveling detailed scattering dynamics between ions and neutrals (*16-27*). In this study, we report product imaging for the charge-transfer collision between spin-orbit selected $Ar^+(^2P_{3/2})$ ions and hydrogen molecules (both *para*- and *ortho*-$H_2$). This reaction is relevant in applications using argon-hydrogen plasmas (*28*); it has been intensively studied in both experiment and theory for half a century, while the detailed charge-transfer dynamics still remain elusive (*29-33*). For the first time, rotational levels of the $H_2^+$ product are clearly resolved at the low collision energy of 0.04 eV and partially resolved at the higher collision energy of 0.22 eV, together with their angular distributions. Full-dimensional quantum scattering and trajectory surface hopping calculations, based on a newly constructed ab initio three-state diabatic potential energy matrix, are performed and the results reproduce the experimental observations reasonably well. This joint experiment-theory study thus provides a unique opportunity to elucidate rotational excitation dynamics in charge-transfer collisions with strong attractive interaction.

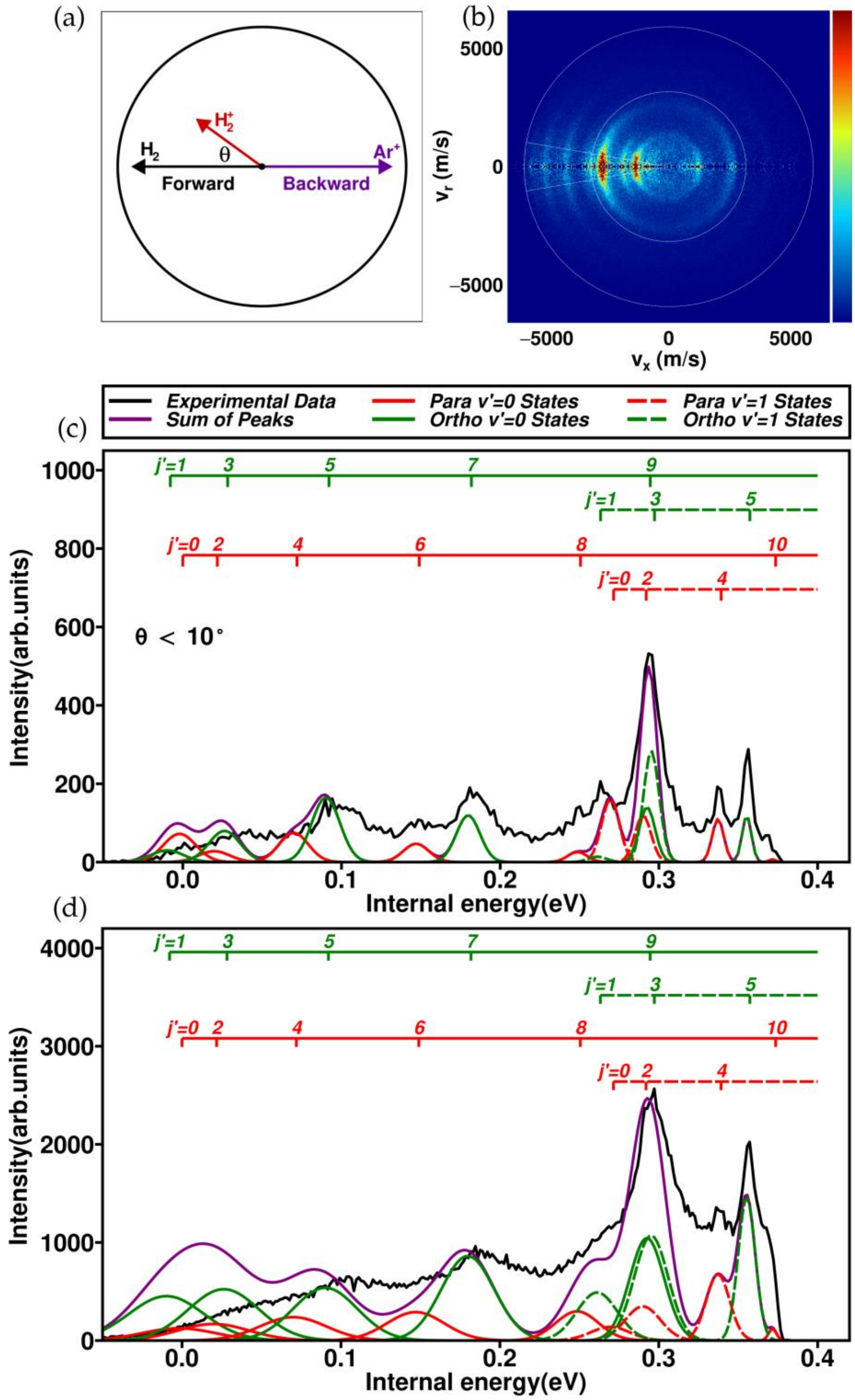

**Figure 1. Experimental differential cross sections for $Ar^+(^2P_{3/2})$ + $H_2(j''=0,1)$ → Ar + $H_2^+(v', j')$ at 0.04 eV of collision energy,** measured with the Innsbruck setup: (a) Newton diagram of the reaction (b) $H_2^+$ product velocity distribution in the center-of-mass frame; the image represents a slice in the collision plane through the three-dimensional velocity distribution; (c) $H_2^+$ product internal energy distributions in the forward scattering region (scattering angle range of 0-10°) (black: experiment, red, green, and purple: quantum calculations (QM) for $j''$=0,1 and the statistical average 25%:75%, markers denote the energies of the product quantum states); (d) $H_2^+$ product internal energy distributions for all scattering angles 0-180°.

**Rotationally resolved product imaging**

The scattering experiments were carried out in a precisely energy-controlled environment on the two ion-molecule crossed-beam setups recently constructed in Innsbruck (*34, 35*) and Beijing (*22, 23*). In both setups, the reactant $Ar^+$ ions were prepared in the spin-orbit level $^2P_{3/2}$ through the resonance enhanced multiphoton ionization method (>97% purity, see Supplementary Information (SI)) (*36*) and crossed with a beam of normal $H_2$ (25% $j''$=0, 75% $j''$=1). The velocity and angular distributions of the product $H_2^+$ ions were measured by a high resolution three-dimensional velocity map ion imaging method. The experimental details are provided in the SI. Figure 1(a) shows the Newton diagram in the center-of-mass frame and Figure 1(b) presents the sliced $H_2^+$ product scattering image for the charge transfer reaction $Ar^+(^2P_{3/2}) + H_2 \rightarrow Ar + H_2^+(v', j')$ at the collision energy of 0.04 eV. The results for 0.22 eV, shown in the SI, display similar scattering dynamics, but with a lower resolution. The white dashed concentric rings represent the kinematic cutoffs for $H_2^+$ in vibrational levels $v'$=0 and 1, respectively, for *para*-$H_2$. As shown in Figure 1(b), the $H_2^+$ products are strongly forward peaked for all the internal energies of $H_2^+$, while a weak peak is also visible in the backward direction. We emphasize that the resolution of rotational levels of the molecular product is unprecedented in an ion-molecule system. The corresponding internal energy distributions are shown in Figures 1(c) for the scattering angle range of 0-10° in the forward scattering direction and 1(d) for the scattering angle range of 0-180° (black solid curves). The most interesting observation here is that the $H_2^+(v'=0)$ product is highly rotationally excited and shows prominent non-statistical characters. More strikingly, these highly rotationally excited products are strongly forward peaked with only a minor peak in the backward region. This observation clearly contradicts the conventional hard-sphere model.

**Quantum dynamics**

To gain insight into the origin of the experimental observations, we constructed a three-state diabatic potential energy matrix, machine-learned from high-level ab initio calculations (details given in the SI). Figure 2(a) shows the three diabatic potentials asymptotically correlated with $Ar^+(^2P_{3/2}) + H_2$, $Ar^+(^2P_{1/2}) + H_2$, and $Ar + H_2^+$, respectively. It is clear that the entrance channels of the exoergic nonadiabatic processes in this system are free of intrinsic barriers. Furthermore, the nonadiabatic charge transfer typically happens at relatively large separations so that the rotational excitation dynamics are dominated by the ground electronic state potential. Quantum scattering calculations were then carried out (for details see the SI), which yielded state-resolved integral and differential cross sections (ICSs and DCSs). As shown in Figure 2(b), the calculated DCS at 0.04 eV is dominated by a strong forward peak with a minor peak in the backward direction, in excellent agreement with the experimental observations as shown in the figure. The $H_2^+(v'=0)$ and $H_2^+(v'=1)$ rotational levels are found to be peaked at $j'=6\sim7$ and $3\sim4$ (Figures 2(c) and 2(d)) ($j'=8\sim9$ and $4\sim5$ at 0.22 eV, respectively, Figure S5 in the SI), respectively, showing prominent non-statistical characters. This is also consistent with the experimental observation. Convoluted with the experimental resolution (see the SI), calculated product internal energy distributions (considering both the initial $H_2(v''=0, j''=0,1)$ states) are displayed in Figures 1(c) and 1(d) (solid red, green, and purple curves), for the forward scattering region of 0-10° and the full scattering angle range of 0-180°, respectively. The agreement between experimental measurements and quantum dynamical calculations is quite satisfactory, especially in the forward scattering region of 0-10°, where also the experimental resolution is higher. This validates the newly constructed diabatic potential energy matrix for capturing the key dynamic features of this ion-molecule system.

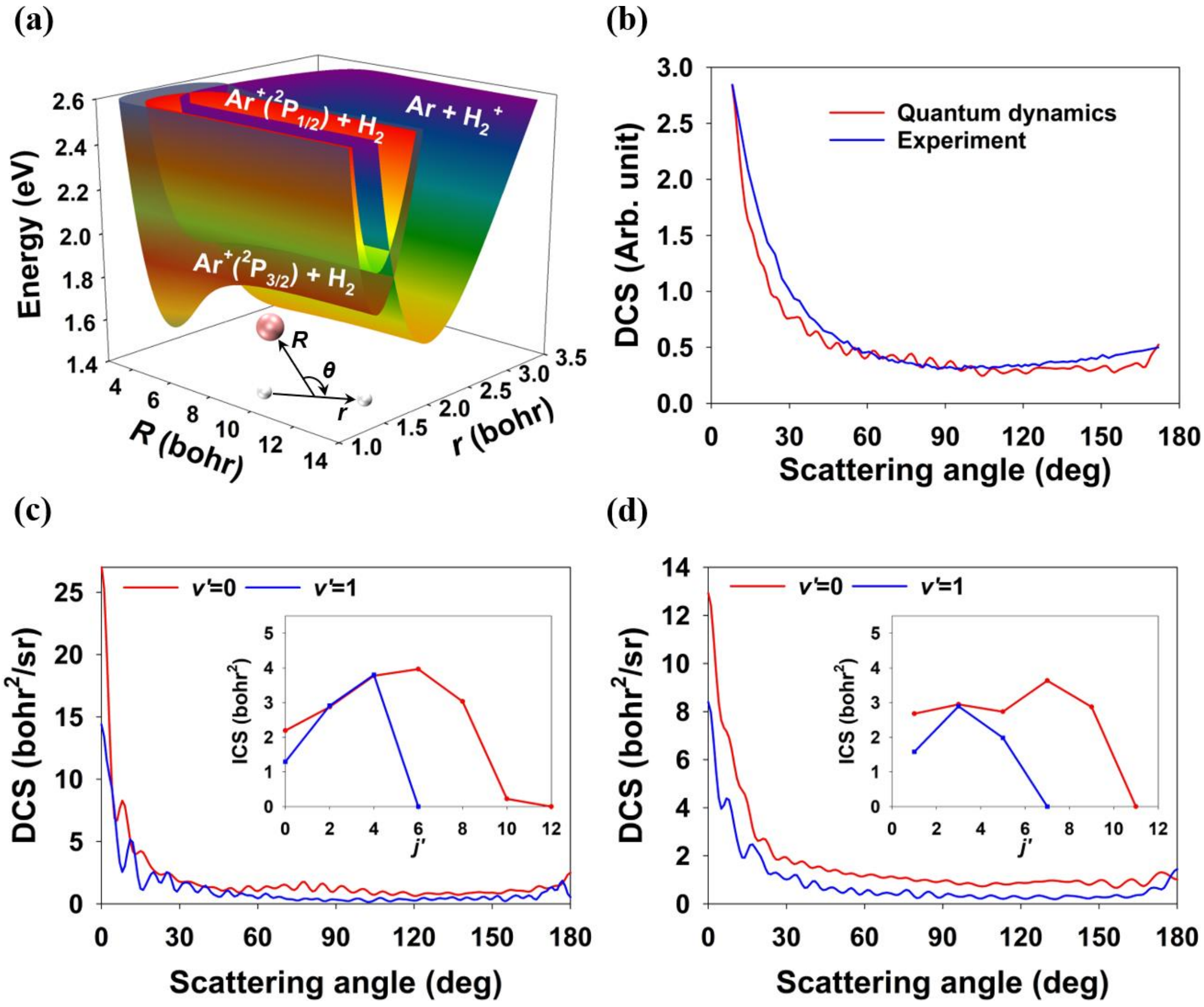

**Figure 2. Diabatic potential energy surfaces (PESs) and quantum scattering results for the charge transfer reaction $Ar^+(^2P_{3/2})$ + $H_2$($v''$=0, $j''$=0, 1) → Ar + $H_2^+$($v'$, $j'$) at the collision energy of 0.04 eV**. (a) diabatic PESs as a function of the Jacobi coordinates $R$ and $r$ with $\theta$ fixed at 90° (see inset); (b) total DCSs at 0.04 eV (red line: quantum dynamics calculations; blue: experiment); product vibrational state resolved DCSs with product rotational distributions shown in the inset for $j''$=0 (c) and $j''$=1 (d).

## Mechanism

As alluded to in the introduction, collisional rotational excitation is often governed by the anisotropy and relative strengths of the short-range repulsive and long-range attractive forces. In the conventional hard-sphere model, rotational excitation is mainly caused by the anisotropic short-range repulsive interactions, while the weak long-range attractive force usually plays only a minor role. The HCGS mechanism only becomes relevant when an exquisite balance exists between the short-range repulsive and long-range attractive forces, in which the latter is sufficiently strong to bend the otherwise sideway scattered rotationally excited products to the forward region. This mechanism therefore typically occurs when the attractive well depth is fractions of the collision energy (*6*).

Here for $Ar^+(^2P_{3/2})$ + $H_2$, a well depth of ~1.4 eV is present in the adiabatic electronic ground state, as shown in Figure S3, which is about 35 times (7 times) of the collision energy 0.04 eV (0.22 eV). This clearly rules out an HCGS mechanism. The agreement between experimental observations and quantum dynamical calculations as presented above provides us with a unique opportunity to investigate the rotational excitation dynamics beyond the hard-sphere model.

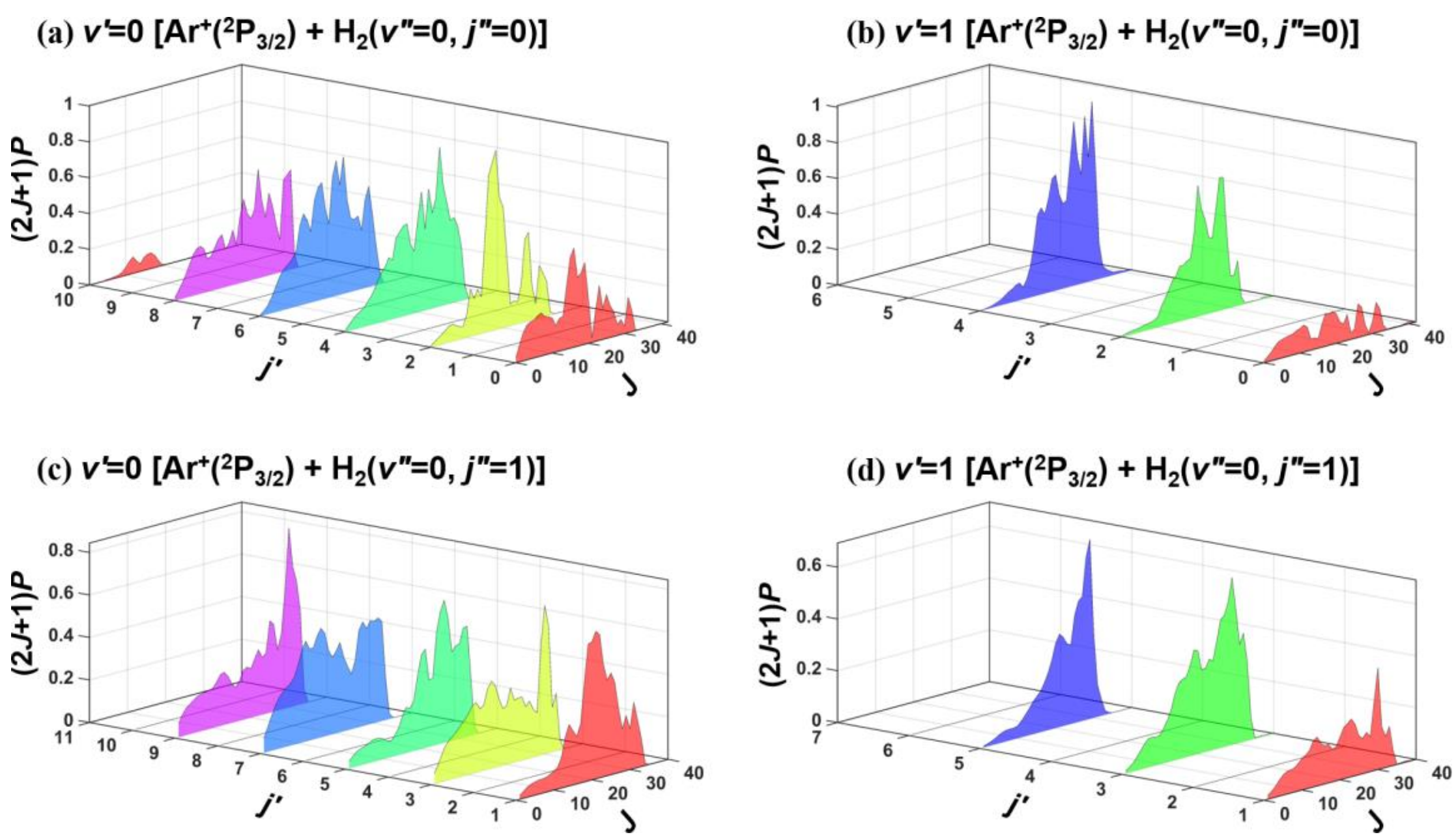

**Figure 3. Quantum opacity functions at the collision energy of 0.04 eV for $H_2(v''=0, j''=0,1)$ excited to various product ro-vibrational levels (*J* denotes the partial wave).**

In Figure 3, quantum opacity functions are presented for excitations from *para/ortho*-$H_2(v''=0, j''=0,1)$ to various final $H_2^+$ product rotational states. Two interesting trends are immediately apparent. First, the higher the rotational excitation of $H_2^+$, the larger the contribution from higher angular momentum *J* partial waves (equivalent to larger impact parameters). This is exactly the opposite of the conventional wisdom based on the hard-sphere model which dictates that collisional rotational excitation is dominated by low *J* partial waves (near-zero impact parameters). Second, all the opacity functions, particularly those corresponding to the highly rotationally excited products, show abrupt cutoffs at $J$=~40 ($J$=~60 at 0.22 eV, Figure S6 in the SI). This is consistent with the range of partial waves contributing to reactions predicted by Langevin capture theory. This observation again suggests that the high product rotational excitation is mainly caused by the long-range "soft" collisions.

To further unravel the physical origin of the rotational excitation process in this system, trajectory surface-hopping (*37*) calculations are performed at the collision energy of 0.04 eV. The scattering angles shown in Figure 4(a) are largely independent of the impact parameter, implying a "complex"-forming mechanism. An exemplary trajectory with forward scattering ($\theta$=8°), an impact parameter of 4 Å and rotational excitation to $j'$=8 is presented in Figure 4(b). After the charge transfer at 349 fs, it is apparent that the two collision partners are trapped in the ground state potential well for a relatively long time (~400 fs) and orbit each other (a movie is provided in the SI). During this orbiting motion, translational energy is gradually converted to the rotational excitation of the product, as shown in the top panel of Figure 4(b). The potential $V$ becomes much lower than that before the charge transfer, as shown in the bottom panel in Figure 4(b). This observation suggests that after the nonadiabatic transition the system is mostly trapped in the ground state well, and the rotational excitation is facilitated largely by the anisotropy of the attractive potential. This picture is completely different from the textbook hard-sphere model, although the orbiting process bears some similarity with the SCBGS except that the trapped $H_2^+$ moiety gains significant rotational angular momentum, due presumably to the anisotropy of the attractive potential well.

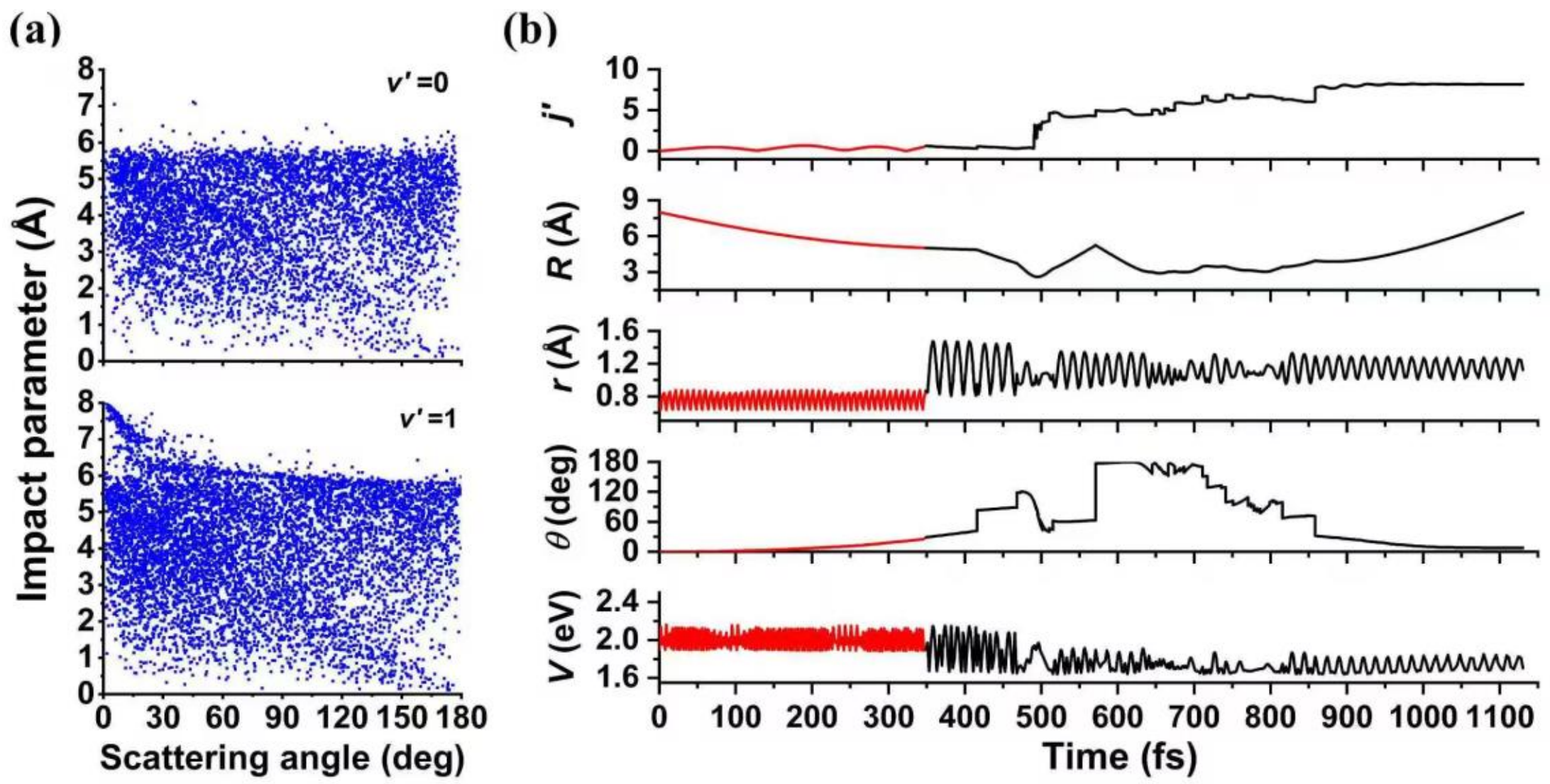

**Figure 4. Trajectory surface hopping results at the collision energy of 0.04 eV.** (a) Correlation between the impact parameter and product scattering angle; (b) evolution of various attributes (rotational level $j'$, Jacobi coordinates $R$, $r$, $\theta$, and potential energy $V$) as a function of time for an exemplary trajectory (red: $Ar^+(^2P_{3/2}) + H_2$; black: $Ar + H_2^+$).

**Concluding remarks**

In this study, we report for the first time imaging of the full state-to-state collision dynamics for a prototypical ion-molecule collision process. Surprisingly, the rotationally excited product is found to scatter mostly in the forward direction, in sharp contrast to the established hard-sphere model. With detailed theoretical analysis, this unexpected phenomenon is attributed to the strong attractive force between the two collision partners. This force traps the system in an orbiting complex, in which the rotational excitation takes place without accessing the short-range repulsive potential. The "soft" dynamical picture is thus completely different from the rotational excitation mechanism that is based on the billiard-ball analogy. The new mechanism is expected to operate in many collisional processes dominated by strong attraction.

**Acknowledgments:** The Innsbruck team acknowledges support from the European Research Council (ERC) under the European Union's Horizon 2020 research and innovation program (grant agreement No. 885479) and by the Austrian Science Fund (FWF) through the Doctoral

Program Atoms, Light, and Molecules, Project No. W1259-N27 [Grant DOI:10.55776/W1259]. The Beijing team is supported by the National Natural Science Foundation of China (Grant Nos. 22525305 and 22373107), Beijing National Laboratory for Molecular Sciences (Grant No. BNLMS-CXXM-202406) and the Innovation Capability Support Program of Shaanxi Province (Grant No. 2023-CX-TD-49). C.X. acknowledges support from the National Natural Science Foundation of China (Grant No. 22573078). H.G. thanks Department of Energy (Grant No. DE-SC0015997). The computation was performed at NWU and at Center for Advanced Research Computing (CARC) at UNM.

**Author contributions**: H.Gao, C.X., H.Guo, and R.W. conceived the project. D.S., J.J., and F.Z. performed the experiment on the Innsbruck setup, supported by T.M., A.K., and R.W.. G.Z. performed the experiment on the Beijing setup, supported by H.Gao. S.H. and C.X. carried out the ab initio calculations and fitting of the coupled potentials, and the quantum dynamics calculations. D.L. and H.Guo performed mechanistic analysis with trajectory surface hopping calculations. All authors are involved in analysis of the data and writing the initial manuscript. H.Gao, C.X., H.Guo, and R.W. wrote the final manuscript.

# Unexpected Collisional Rotational Excitation via Long-Range Capture and Orbiting – Supplementary Information –

Dasarath Swaraj,[a,†] Guodong Zhang,[a,‡,¶] Siting Hou,[a,§] Dandan Lu,[a,∥] Jerin Judy,[†] Fabio Zappa,[†] Tim Michaelsen,[†] Arnab Khan,[†] Hua Guo,[*,⊥] Changjian Xie,[*,§] Hong Gao,[*,‡,¶] and Roland Wester[*,†]

†*Institut für Ionenphysik und Angewandte Physik, Universität Innsbruck, Technikerstraße 25, 6020 Innsbruck, Austria*

‡*Beijing National Laboratory for Molecular Sciences (BNLMS), Institute of Chemistry, Chinese Academy of Sciences, Beijing 100190, China*

¶*University of Chinese Academy of Sciences, Beijing 100049, China*

§*Institute of Modern Physics, Shaanxi Key Laboratory for Theoretical Physics Frontiers, Northwest University, Xi'an, Shaanxi 710127, China*

∥*State Key Laboratory of High Temperature Gas Dynamics, Institute of Mechanics, Chinese Academy of Sciences, 100190 Beijing, China*

⊥*Department of Chemistry and Chemical Biology, Center for Computational Chemistry, University of New Mexico, Albuquerque, New Mexico 87131, USA*

E-mail: hguo@unm.edu; chjxie@nwu.edu.cn; honggao2017@iccas.ac.cn; roland.wester@uibk.ac.at

[a]These authors contributed equally to this work.

# Energy Level Diagram

Fig S1 shows the energy level diagram for the studied reaction

$$\mathrm{Ar}^+(^2P_{3/2}) + \mathrm{H}_2(v'' = 0, j'' = 0, 1) \rightarrow \mathrm{H}_2^+(v', j') + \mathrm{Ar}(^1S_0) \tag{S1}$$

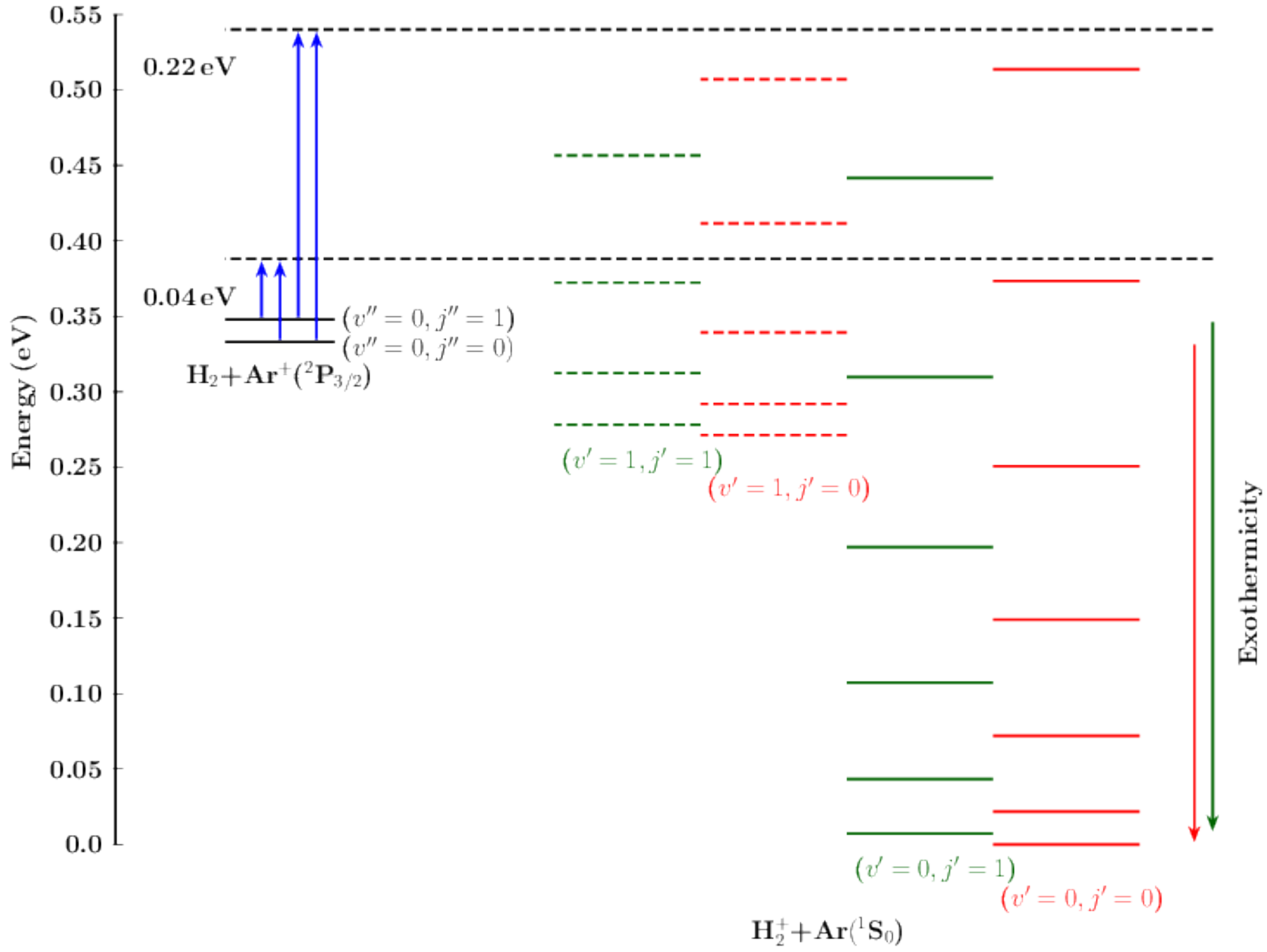

Figure S1: Energy level diagram for the first relevant rotational states of $H_2$ (black) and $H_2^+$ (red and green). The arrows show exothermicities for reactions of para (red )and ortho (green) $H_2$, respectively. The black dashed lines represent the two studied collision energies.

## Internal Energy Distribution

In the center of mass, the total kinetic energy of the two products after the reaction will be given by the energy difference between the initial and final states, as depicted in Figure S1, plus the collision energy, $E_{col}$:

$$E_{kin} = E_{col} + E(\nu'', j'') - E'(\nu', j') \tag{S2}$$

The exothermicity of the reaction is given by the difference in energy between the ground states of reactants and products, i. e. for *para*-$H_2$.

$$E_{ex} = E(0,0) - E'(0,0) \approx 0.33 \text{ eV} \tag{S3}$$

We define the kinematic cut-off as the sum of collision energy and exothermicity. This establishes the maximum for the measured product kinetic energies or the zero point of the measured internal energy.

$$E_{cut} = E_{col} + E_{ex} \tag{S4}$$

The experimental measurements extract the velocity vectors of the ionic products after the collision. From the velocities, the total product kinetic energies are calculated, and the internal energy of the product molecules is obtained using the expression

$$E_{int} = E_{cut} - E_{kin} \tag{S5}$$

Replacing S2, S3 and S4 in S5, we obtain all possible peak positions in our internal energy distributions, given by

$$E_{int} = E(0,0) - E'(0,0) - E(\nu'', j'') + E'(\nu', j') \tag{S6}$$

The total hydrogen nuclear spin is expected to be conserved in the charge transfer reaction with $Ar^+$. As a consequence, transitions involving odd-even or even-odd rotational quantum numbers are forbidden and the expression for the allowed distribution mean positions has to obey the condition:

$$E_{\text{int}}(\nu', j') = \begin{cases} E'(\nu', j') - E'(0,0), & \text{if } j' \text{ is even,} \\ \\ E'(\nu', j') - E'(0,0) - \big[E(0,1) - E(0,0)\big], & \text{if } j' \text{ is odd.} \end{cases} \tag{S7}$$

The markers in Figure 1 of the main manuscript and in Figure S4 are calculated accordingly.

# Experimental Methods

## Innsbruck Setup

The Innsbruck data have been obtained with a new crossed-beam imaging spectrometer, which was constructed to produce well-defined initial ion and neutral molecular beams for very low-energy collisions.[34] In this setup, the ion beam can be intersected by one out of two neutral beams between the two lowest electrodes of a velocity map imaging (VMI) spectrometer.[38] One neutral beam reaches the interaction region at an angle of +45° with respect to the ion beam, the other one at an angle of -45°. For the data presented here, only one of the two neutral beams was used. However, we confirmed that with the other neutral beam identical imaging data could be obtained.

The reactant neutral hydrogen molecules (purity 5.0) are expanded into vacuum through a pulsed Even-Lavie valve, using a backing pressure of 50 bar.[39] This produces a supersonic beam, which reaches the interaction region through a 3 mm skimmer. The valve is mounted on an XYZ manipulator to align with the skimmer to obtain the maximum neutral beam density at the reaction region.

The $Ar^+$ ions are generated from argon gas (purity 6.0), expanded to ultrahigh vacuum using a home-built piezo valve operated at 100 Hz. Following the supersonic expansion, argon atoms are laser-ionized using a 3+1 Resonance Enhanced Multi-Photon Ionization (REMPI) scheme at 314.466 nm to produce $Ar^+$ in the spin-orbit ground state.[36] The laser wavelength is obtained by doubling the output of a pulsed dye laser with dye 4-(dicyanomethylene)-2-methyl-6-(4-dimethylaminostyryl)-4H-pyran, which is pumped from the frequency-doubled output of a Nd:YAG pulsed laser at 532 nm.

The ion packet is formed between the first two electrodes of a set of nine electrodes, to which static voltages are supplied.[34] With these voltages the ions are guided to the reaction region. Before entering the VMI spectrometer, a set of ten electrostatic lenses is used to decelerate and focus the ions onto the neutral $H_2$ reactant beam. The collision energy is controlled by tuning the energy of the reactant ions, which is defined as the difference between the average voltages provided by the first two electrodes for ion acceleration and the offset voltage applied to the VMI spectrometer.

The VMI spectrometer consists of seven circular plates, which are 24 mm apart. All except the lowest one, the repeller, include a shielding cylinder to suppress the perturbation by stray electric fields in the volume below the electrodes. Except for repeller, all electrodes contain a central circular opening, with a diameter of 30 mm in the second electrode, the extractor, and a diameter of 40 mm in the other five electrodes. Ions and neutral molecules collide in the center of the volume defined by the lowest two VMI electrodes, the repeller and extractor plates. The first six electrodes, all except the final grounded one, are switched to appropriate high voltages after the ion and neutral beam bunches collide. The rise time of the plate voltages is approximately 40 ns. Product ions are detected using a position-sensitive scheme, consisting of a micro-channel plate, a phosphor screen, and a CMOS camera. The impact positions of the ions on the detector are converted to velocity vectors in the horizontal plane. The vertical component of the velocity is obtained from the measurement of the arrival times to the detector, which is measured using a photo-multiplier tube placed above the phosphor screen, on the side of the camera.

The velocity distributions of the reactant beams are characterized with the VMI spectrometer using the same ion optics settings that are also used to record product reactive scattering. For the ion energy distribution a mean value of 1.3 eV and a FWHM of 0.07 eV was measured, which provides a mean relative energy for the collision with the $H_2$ reactants of 0.04 eV.

To measure the velocity distribution of the neutral reactant, $H_2$ molecules from the supersonic beam are ionized with multiple photons from a pulsed laser at 266 nm. This wavelength is obtained by doubling twice the output of an Nd:YAG pulsed laser. The velocities of the two neutral beams were determined to be 2600 m/s, with a FWHM of $\approx$ 300 m/s in both cases. This velocity spread is caused by the actual velocity spread of the beam combined with a finite broadening due to the ion recoil in the ionization process.

## Beijing Setup

The state-selected ion-molecule crossed-beam apparatus in Beijing has been described in detail previously.[22,23] The resonance-enhanced multiphoton ionization (REMPI) method was used to produce the $Ar^+$ ions in the spin-orbit ground $^2P_{3/2}$ state with a purity better than 97%.[36] The ultraviolet (UV) laser used for the REMPI process was generated by doubling the fundamental output of a dye laser (LiopTec, LIOPSTAR-E) pumped by the second-harmonic output of a 10-Hz Nd:YAG laser (Beamtech, Nimma-900). The UV wavelength was set at 314.466 nm with a pulse energy of 0.8 mJ and focused into the photoionization region by a planoconvex lens with a focal length of 150 mm. A pulsed pure atomic beam of Ar was generated by a general valve (Parker, Series 9) with a stagnation pressure of 4 bar and photoionized by a (3 + 1) REMPI process.[36] The generated $Ar^+$ ion beam was then accelerated, focused and finally decelerated to the target kinetic energy before arriving at the reaction center, by applying proper voltages to the corresponding electrodes.[23] In the center of the three-dimensional velocity map imaging (3D-VMI) setup (the reaction center), the $Ar^+$ ion beam crossed with the pulsed supersonic $H_2$ beam at 90°, where the $H_2$ beam was produced by an Even-Lavie valve with a stagnation pressure of 15 bar. A (3+1) REMPI spectrum of the supersonic $H_2$ beam was collected,[40] which showed that most $H_2$ molecules are cooled down to the lowest two rotational levels $j'' = 0$ (*para*-$H_2$) and $j'' = 1$ (*ortho*-$H_2$), and the populations of higher rotational levels are negligible. The kinetic energies of the $Ar^+$ and $H_2$ beams were directly measured by the 3D-VMI system and had typical spreads of 170 meV at the total kinetic energy of 3.66 eV and 13 meV at the total kinetic energy of 55 meV along their flying directions, respectively. To suppress the space charge effect and maintain a reasonable signal-to-noise level, ≈150 $Ar^+$ ions are produced in each pulse. For the scattering product image at the collision energy of 0.22 eV, 100,000 product ions were collected to obtain sufficient statistics.

# Methods and analysis

The methods and analysis described here correspond to the dynamics results at the collision energy of 0.04 eV

## Kinematic broadening

The main source for the product velocity uncertainty in a crossed-beam experiment is the finite velocity resolution of reactant beams, both in the parallel and perpendicular directions. The kinematic resolution is determined using Gaussian error propagation for the product ions' velocity magnitude and angle in the center of mass as a function of resolution parameters obtained from the beam characterization measurements and the scattering angle (see Experimental Methods, above).[34, 41] For the scattering data obtained with the Innsbruck setup, the kinematic velocity resolution for products flying in the forward direction is obtained from the Gaussian error propagation to be $\Delta v_{\mathrm{H}_2^+} \approx 80\,\mathrm{m/s}$.

## Resolution of the VMI spectrometer

The second source for the product velocity uncertainty is the finite resolution of the VMI spectrometer. This resolution is caused by an imperfect velocity map imaging, i. e. the extension of the interaction volume is not perfectly focused onto the imaging detector for ions with the same velocity and by the finite position resolution of the imaging detector. For the Innsbruck VMI spectrometer the resolution due to imperfect mapping is estimated to be approximately 10 m/s, obtained from trajectory simulations of ions passed through the spectrometer. The detector position resolution is estimated to be around $30\,\mu m$, which corresponds to the spacing between the pores of the microchannel plate. This translates into a velocity resolution of about 15 m/s.

## Broadening in Internal Energy

To compare the experimental internal energy distributions with the theoretical state-to-state cross sections, shown in Figure 1 of the main manuscript, the product states of the theoretical calculations have to be convoluted with the experimental energy resolution. To determine the internal energy broadening, a transformation of the spectrometer velocity resolution is required.

The product velocity uncertainty $\Delta v_{\mathrm{H}_2^+}$ of the measured $\mathrm{H}_2^+$ ions propagates to an uncertainty of the total product kinetic energy $E_{\mathrm{kin}}$, which depends on the square of the relative product velocity $v_{\mathrm{rel}} = \frac{m_{\mathrm{Ar}}+m_{\mathrm{H}_2^+}}{m_{\mathrm{Ar}}} v_{\mathrm{H}_2^+}$. Specifically, it is given by

$$\Delta E_{\mathrm{kin}} = \sqrt{2\mu E_{\mathrm{kin}}}\,\frac{m_{\mathrm{Ar}} + m_{\mathrm{H}_2^+}}{m_{\mathrm{Ar}}}\,\Delta v_{\mathrm{H}_2^+} \tag{S8}$$

$m_{\mathrm{Ar}}$ and $m_{\mathrm{H}_2^+}$ are the masses of argon atoms and hydrogen molecular ions, respectively, and $\mu$ is the reduced mass.

To incorporate the energy uncertainty into the presented figures of the internal energies, Gaussian distributions were generated for each product quantum state. Given the above mentioned scaling, the standard deviation $\sigma_{(\nu',j')}$ of each Gaussian distribution depends on the internal quantum state in the form

$$\sigma_{(\nu',j')} = \sqrt{2\mu(E_{\mathrm{cut}} - E_{\mathrm{int}}^{(\nu',j')})}\frac{m_{\mathrm{Ar}} + m_{\mathrm{H}_2^+}}{m_{\mathrm{Ar}}}\Delta v_{\mathrm{H}_2^+} \tag{S9}$$

The energy of the kinematic cutoff, $E_{\mathrm{cut}}$, is defined in section on the internal energy distribution, above.

According to the expression for $\sigma_{(\nu',j')}$, lower energy product quantum states, corresponding to faster product velocities, are broadened stronger than higher energy quantum states. This is in excellent agreement with experiment (see Figure 1 of the main manuscript). For quantum states approaching the highest accessible energy, corresponding to the near-zero product velocities in the center-of-mass frame, the expected standard deviation vanishes.

For the internal energy, also the uncertainty of the kinematic cutoff matters, which is caused by the fluctuations of the relative collision energy between the two reactants. This uncertainty is calculated to be $\sigma_E = 2\,\mathrm{meV}$. To account for this, the simulated distribution is folded with a second Gaussian distribution with $\sigma_E$ as a constant width in energy. Quadratically adding the two standard deviations $\sigma_{(\nu',j')}$ and $\sigma_E$, one obtains

$$\sigma^*_{(\nu',j')} = \sqrt{\Delta v^2 \left(E_{\text{cut}} - E_{\text{int}}^{(\nu',j')}\right) + \sigma_E^2} \tag{S10}$$

With this standard deviation, a Gaussian distribution is generated for each quantum state

$$P^{(\nu',j')}(E_{\text{int}}) = \frac{1}{\sqrt{2\pi}\sigma^*_{(\nu',j')}} \exp\left(-\frac{1}{2}\left(\frac{E_{\text{int}} - E^{(\nu',j')}}{\sigma^*_{(\nu',j')}}\right)^2\right). \tag{S11}$$

Finally, the internal energy distributions obtained from the simulations, shown in Figure 1 of the main manuscript, are given by the sum over all accessible product quantum states

$$P'(E_{\text{int}}) = \sum_{\nu',j'} A_{(\nu',j')} P^{(\nu',j')}(E_{\text{int}}) \tag{S12}$$

The peak amplitudes $A_{(\nu',j')}$ of the distributions for each rovibrational state are taken from the theoretically calculated state-to-state cross sections.

# Theory

## *Ab initio* calculations

The electronic structure calculations for the $ArH_2^+$ system were carried out using the multi-reference configuration interaction (MRCI) method with the Davidson correction (MRCI+Q).[42–44] The augmented correlation-consistent polarized valence quadruple-zeta (aug-cc-pVQZ)[45] basis set was used for all atoms. The state-averaged complete active space self-consistent field (SA-CASSCF) calculations were first carried out with the active space (9e, 6o), which includes the lowest three $A'$ and one $A''$ doublet states with equal weights. Based on the SA-CASSCF orbitals, the electronic energies of the $1A'$, $2A'$, and $3A'$ states were simultaneously computed at the MRCI+Q level.

The calculations of the derivative couplings were performed by the central-difference algorithm with a small displacement 0.0001 bohr. To construct the coupled PESs for the charge transfer reaction $Ar^+ + H_2 \rightarrow Ar + H_2^+$, the *ab initio* points were sampled in the geometric configuration within the ranges of [0.8, 4.0] bohr, [2.0, 50.0] bohr, and [0, 90] degree in the Jacobi coordinates $(r, R, \theta)$, in which $r$ is the H-H bond length, $R$ is the distance between Ar and the center of mass of $H_2$, and $\theta$ denotes the angle between the vectors $R$ and $r$, respectively. Simultaneously, numerous points were sampled to cover the $ArH^+$ + H channel with the ranges of [1.8, 3.6] bohr and [0.8, 30.0] bohr for the Ar-H and H-H bond distances, respectively, and [0, 180] degree for the inter-bond angle between the Ar-H and H-H bond lengths. All the *ab initio* calculations reported in this work were performed using MOLPRO.[46,47]

## Diabatization of $1A'$, $2A'$, and $3A'$ states

There are four doublet electronic states ($1A'$, $2A'$, $3A'$, and $1A''$) correlated to the $Ar^+$ + $H_2$ and Ar + $H_2^+$ channels. At the $H_2$ equilibrium geometry with a bond length of 1.40 bohr, the $1A'$, $2A'$, and $1A''$ states are correlated to the $Ar^+ + H_2$ asymptote, while at the $H_2^+$ equilibrium geometry with a bond length of 1.95 bohr, the Ar + $H_2^+$ channel becomes the lowest ($1A'$) and the $2A'$, $3A'$, and $1A''$ states are correlated to the $Ar^+$ +

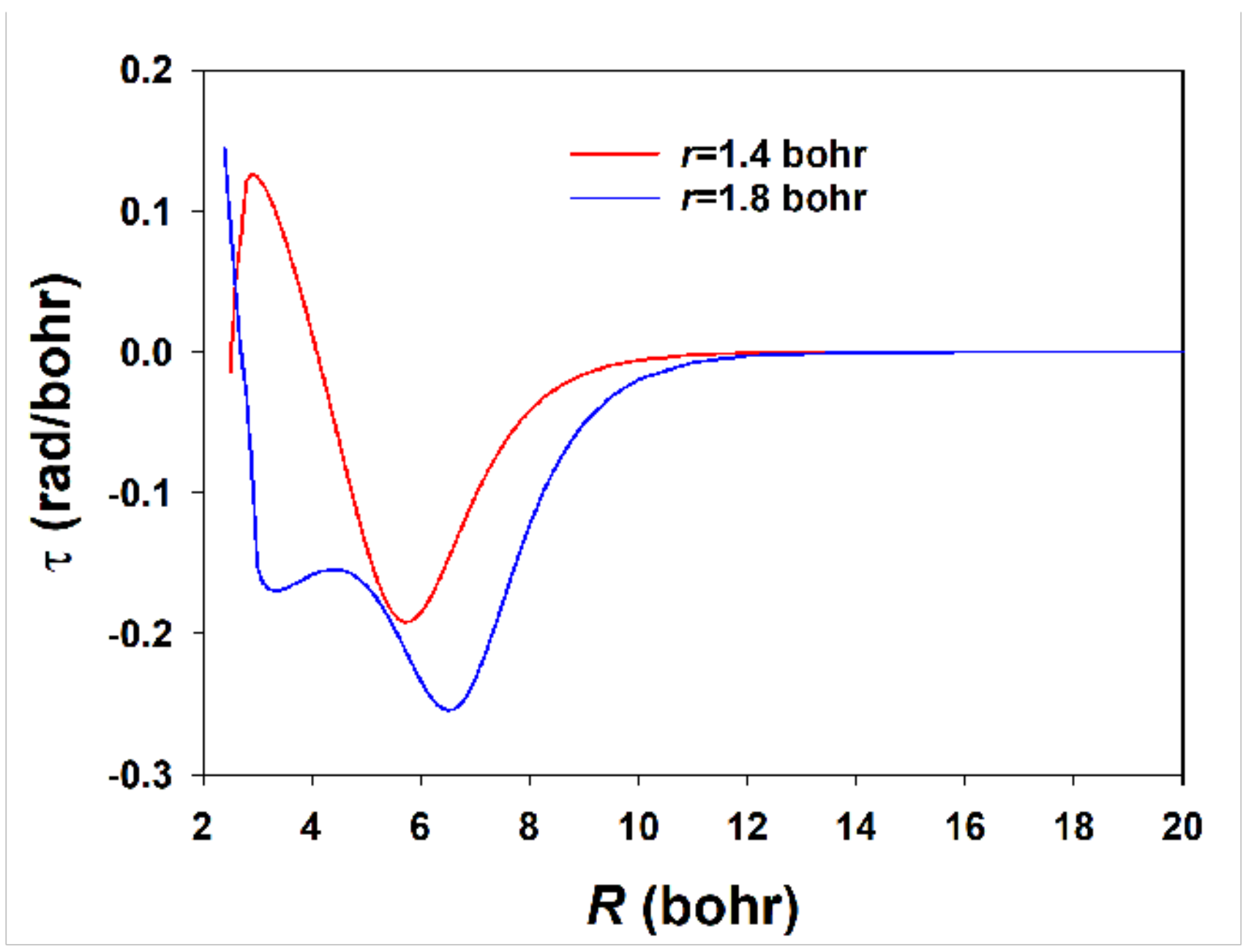

Figure S2: *Ab initio* derivative couplings between the $1A_1$ and $2A_1$ states for r=1.4 and 1.8 bohr at $C_{2v}$ symmetry ($\theta$=90 deg).

$H_2$ asymptote. Because there is no coupling between $A'$ and $A''$ states, the $1A''$ state was not included in the construction of the diabatic potential energy matrix (DPEM). At $C_{2v}$ symmetry, the couplings among the three $A'$ states (in $C_s$ symmetry) become simpler since they belong to $1A_1$, $2A_1$, and $1B_2$, respectively. Only the couplings between two $A_1$ states are nonzero and need to be considered.

In this work, the diabatization in $C_{2v}$ symmetry between $1A_1$ and $2A_1$ was achieved using a mixing angle determined by the integral of the derivative coupling t along the $R$ coordinate. Figure S2 shows the derivative couplings at $C_{2v}$ symmetry ($\theta$=90 deg) along the $R$ coordinate for the H-H bond length fixed at 1.4 and 1.8 bohr.

Based on the derivative coupling, the mixing angle $\alpha$ between the two $A_1$ states can be computed by:[48]

$$\alpha = \int_{R_0}^{R_1} \tau(R) dR \tag{S13}$$

where $R_0$ was chosen at 30 bohr in the calculations and $R_1$ denotes the point that needs to be transformed into diabatic energies from adiabatic energies using the following expressions:

$$V_{11} = E_1 cos^2\alpha + E_2 sin^2\alpha \tag{S14}$$

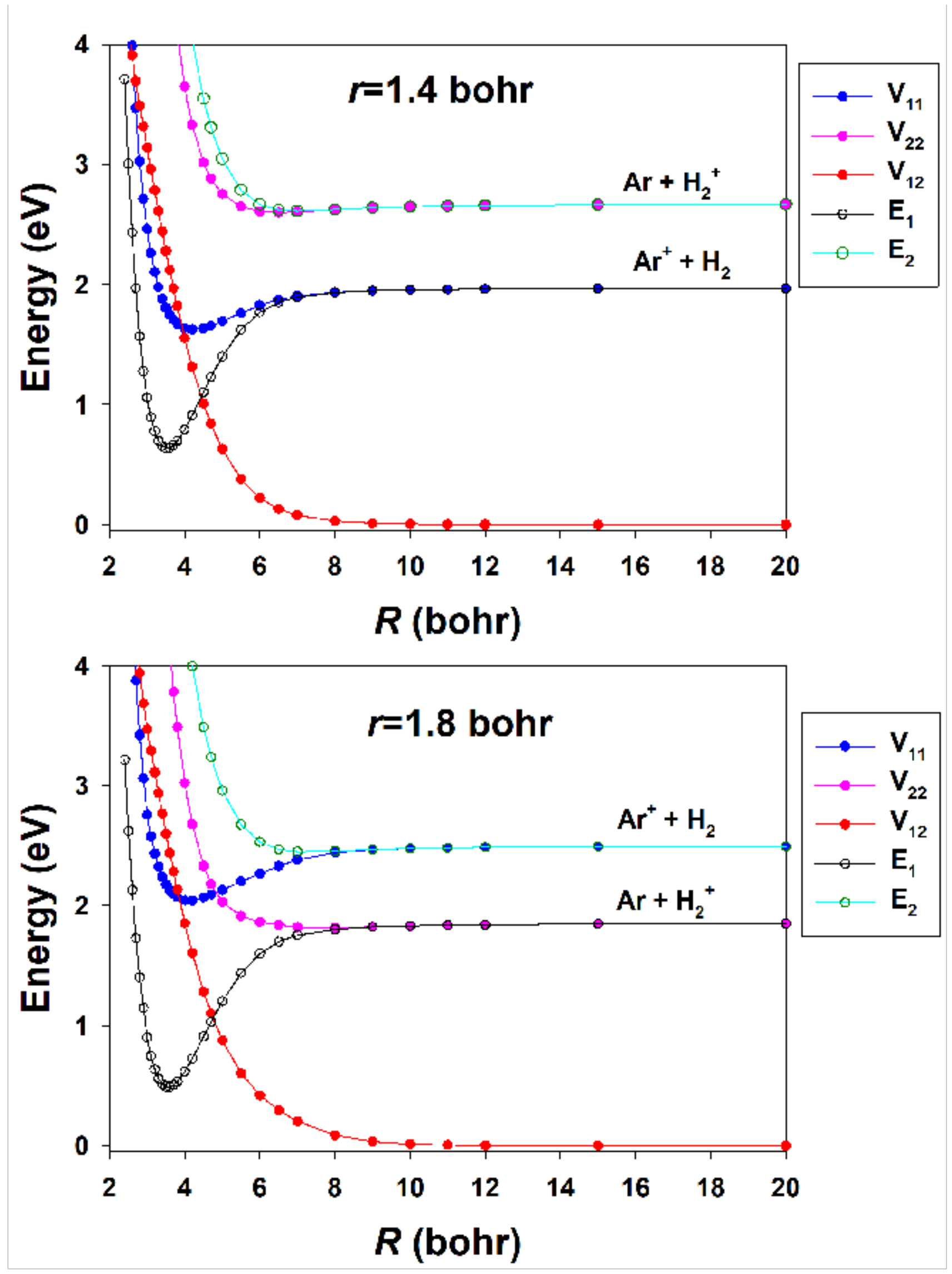

Figure S3: Diabatic and adiabatic potential energy curves for the $1A_1$ and $2A_1$ states at $r$ =1.4 and 1.8 bohr at $C_{2v}$ symmetry ($\theta$=90 deg).

$$V_{22} = E_1 sin^2\alpha + E_2 cos^2\alpha \tag{S15}$$

$$V_{12} = 0.5(E_2 - E_1)sin2\alpha \tag{S16}$$

in which $E_1$ and $E_1$ are adiabatic potential energies and $V_{11}$, $V_{22}$, and $V_{12}$ are diagonal and off-diagonal elements of the DPEM for two $A_1$ states. Figure S3 shows the diabatic and adiabatic potential energy curves for the H–H bond length fixed at 1.4 and 1.8 bohr.

For the linear geometric configurations, the three $A'$ states also correspond to the $1A_1$, $2A_1$, and $1B_2$ states at $C_{2v}$ symmetry. Thus, the diabatization of the $1A_1$ and $2A_1$ states was analogously accomplished.

## Neural network representation of DPEM

In this work, elements of the DPEM are represented by multilayer feedforward neural networks (NNs)[49] consisting of an input layer, hidden layers, and an output layer. For a given layer ($i$+1), the form of the $k$th neuron can be written as

$$y_k^{i+1} = f_{i+1}\left(\sum_{j=1}^{N_i} w_{k,j}^{i+1} y_k^i + b_k^{i+1}\right) \tag{S17}$$

in which $N_i$ is the number of neurons in the $i$th layer, $w_{k,j}^{i+1}$ are the weights connecting the $j$th neuron in the $i$th layer and the $k$th neuron in the ($i$+1)th layer, and are the biases of the $k$th neuron in the ($i$+1)th layer. $f_{i+1}$ are the transfer functions for the ($i$+1)th layer. The input vector includes three variables of the coordinates, which are chosen as the following permutation invariant polynomials:[50]

$$G_1 = 0.5\left(\frac{1}{r_{13}} + \frac{1}{r_{23}}\right) \tag{S18}$$

$$G_2 = \frac{1}{r_{13}r_{23}} \tag{S19}$$

$$G_3 = \frac{1}{r_{12}} \tag{S20}$$

in which $r_{ij}$ is the bond length of two atoms, 1 and 2 denote the two H atoms, and 3 stands for Ar atom. The units for three radial coordinates are Å. Without the spin-orbit couplings, the 3-state coupled DPEM can be expressed as:

$$\mathbf{V} = \begin{pmatrix} V_{11} & V_{12} & V_{13} \\ V_{21} & V_{22} & V_{23} \\ V_{31} & V_{32} & V_{33} \end{pmatrix} \tag{S21}$$

in which $V_{11}$, $V_{22}$ and $V_{33}$ denote the $1A_1$, $2A_1$, and $1B_2$ states, respectively. The off-diagonal elements stand for the diabatic coupling terms. Each element of the DPEM is represented by an individual NN function, which is smooth as a function of the three nuclear coordinates. Specifically, a characteristic factor $\lambda = (r_{13} - r_{23})\sin\theta$ was imposed on the NN function of the coupling terms $V_{13}$ and $V_{23}$. The usage of $r_{13} - r_{23}$ is due to the asymmetric couplings with respect to the permutation of the two H atoms[51,52] and $\sin\theta$ is used to maintain the symmetry of the couplings at the linear geometries. For the diabatic coupling term $V_{12}$, there is no additional symmetry constraint since two states carry the same symmetry. In addition, the diabatic coupling should approach zero in the $Ar^+$ + $H_2$ and Ar + $H_2^+$ channels since the interactions between atoms and diatoms are negligible in the asymptote and the electronic states of the species are well separated. Therefore, a constrained damping factor was imposed on all the diabatic couplings with the following form:[51]

$$d = 0.5\left(\tanh\left(6(7.5 - r_m)\right) + 1\right) \tag{S22}$$

in which $r_m$ is the maximal value of the two Ar-H bond lengths (in Å). In the fitting procedure, the NN parameters were determined by minimizing the residual sum of squares:[49]

$$\begin{aligned} S = & \frac{1}{N}\sum_{i=1}^{N}\sum_{j=1}^{3} w_j\left(E_{i,j}^{\text{fit}} - E_{i,j}\right)^2 + \frac{1}{M}\sum_{j=1}^{M}\sum_{i=1}^{3} w_j\left(V_{ii,j}^{\text{fit}} - V_{ii,j}\right)^2 \\ & + \beta\sum_{i=1}^{L}\sum_{n=1}^{2}\sum_{n'=n+1}^{3}\left(\tilde{f}_{\text{fit},i}^{\,n,n'} - \tilde{f}_i^{\,n,n'}\right)^2 + 0.5t\left\|\mathbf{p}\right\|^2 \end{aligned} \tag{S23}$$

where $E$ are the adiabatic energies and $N$ is the number of adiabatic energy points, $w$ are the energy dependent weights. $V_{ii}$ are diabatic energies that obtained at $C_{2v}$ symmetry for different H-H bond lengths and $M$ is the number of the diabatic energy points. $\tilde{f}_i^{\,n,n'}$ are the energy scaled derivative couplings, defined as $\tilde{f}_i^{\,n,n'}=f_i^{\,n,n'}\left(E_{i,n'} - E_{i,n}\right)$, where $f_i^{\,n,n'}$ are the derivative couplings along the $R$ direction and their absolute values were used. $\beta$ is a scale factor that was set as 0.05 in the calculations to obtain a good fit, and $L$ is the number of the energy scaled derivative coupling data. The $0.5t\left\|\mathbf{p}\right\|^2$term is the regularization term that is used to avoid overfitting. $\mathbf{p}$ denotes all the optimized parameters (weights and biases) of the NN functions and $t$ is a small number ($10^{-5}$ in the calculations). The Levenberg-Marquardt algorithm[53] was used to iteratively optimize the parameters and the Jacobian matrix was numerically calculated by the central-difference algorithm with the displacements $10^{-5}$.

In the NN fitting of DPEM, 10 neurons in each of the two hidden layers for all the elements of DPEM as expressed in Eq. (S21) were used, resulting in 966 parameters. 21642, 1463, and 866 geometries were included for the adiabatic energy, diabatic energy, and energy scaled derivative coupling data, respectively. Because the collision energies in experiment are not very high (0.04 and 0.22 eV), points with energies below 4 eV relative to the minimum on the ground adiabatic electronic state carried a significantly larger weight in the fitting, as shown in Table S1. The root mean square errors (RMSEs) of $E_1$, $E_2$, and $E_3$ below 4 eV are 20.1, 9.8, and 6.9 meV, respectively, while for the diabatic energies $V_{11}$, $V_{22}$, $V_{33}$, and $V_{12}$, the RMSEs are 14.5, 13.8, 7.3, and 18.8 meV, respectively. The RMSEs of the energy scaled derivative couplings $\tilde{f}^{12}$, $\tilde{f}^{13}$ and $\tilde{f}^{23}$ are 0.0077, 0.033, and 0.0049 eV/bohr, respectively. The representation of the ab initio data by the NN DPEM is considered to be excellent.

Finally, the spin-orbit coupling was further considered as a constant for the spin-orbit corrected DPEM, which is given by[54]

$$\mathbf{V} = \begin{pmatrix} V_{11} - A & V_{12} & V_{13} \\ V_{21} & V_{22} - B & V_{23} \\ V_{31} & V_{32} & V_{33} + 2A \end{pmatrix} \tag{S24}$$

where $A$ denotes the spin–orbit splitting term and was set as 0.0591 eV to match the experimental derived 0.178 eV energy splitting between $Ar^+(^2P_{3/2})$ and $Ar^+(^2P_{1/2})$. $B$ was set as 0.1017 eV to match the experimental exothermicity (0.334 eV) of the charge transfer reaction $Ar^+(^2P_{3/2}) + H_2 \rightarrow Ar + H_2^+$. The diabatic states 1, 2, and 3 represented in Eq. S24 denote the $Ar^+(^2P_{3/2}) + H_2$, $Ar + H_2^+$, and $Ar^+(^2P_{1/2}) + H_2$ channels.

## Quantum dynamics

In this work, the Chebyshev real wave packet method[55–57] was used to extract state-to-state scattering amplitudes of the charge transfer reaction $Ar^+ + H_2(v'', j'') \rightarrow Ar + {H_2}^+(v', j')$. Since the details of the scattering theory can be found in earlier papers,[58–60] only a brief description is given here. The Hamiltonian can be written as

$$\mathbf{H} = \hat{T} \begin{pmatrix} 1 & 0 & 0 \\ 0 & 1 & 0 \\ 0 & 0 & 1 \end{pmatrix} + \mathbf{V} \tag{S25}$$

where $\hat{T}$ is the kinetic energy operator in the $Ar^+$ + $H_2$ Jacobi coordinates $(R, r, \gamma)$ in atomic units:

$$\hat{T} = -\frac{1}{2\mu_R}\frac{\partial^2}{\partial R^2} - \frac{1}{2\mu_r}\frac{\partial^2}{\partial r^2} + \frac{(\hat{J}-\hat{j})^2}{2\mu_R R^2} + \frac{\hat{j}^2}{2\mu_r r^2} \tag{S26}$$

where $\mu_r = \frac{m_H}{2}$ and $\mu_R = \frac{m_{Ar}\, m_{H_2}}{m_{Ar}+m_{H_2}}$ are the corresponding reduced masses respectively, $J$ and $j$ are the total angular momentum and the diatom rotational angular momentum. $\mathbf{V}$ in Eq. (S25) is the 3-state DPEM as described above.

In the diabatic representation, the nuclear wavefunctions associated with the three electronic states are given by:

$$\Psi = \begin{pmatrix} \psi^1 \\ \psi^2 \\ \psi^3 \end{pmatrix} \tag{S27}$$

where '1', '2', and '3' denote the $Ar^+(^2P_{3/2})$ + $H_2$, Ar + $H_2{}^+$, and $Ar^+(^2P_{1/2})$ + $H_2$ channels, respectively. Each nuclear wavefunction is discretized in a mixed representation,[58–60] consisting of a direct product discrete variable representation (DVR)[61] with the sine function for the two radial degrees of freedom and a finite basis representation (FBR) with the associated Legendre polynomial for the angular degree of freedom.

The initial wave packet was chosen as the product of the $H_2(v'' = 0, j'' = 0, 1)$ rovibrational eigenfunctions and a real Gaussian wave packet in the translational degree of freedom $(R)$ in channel 1, corresponding to the $Ar^+(^2P_{3/2})$ + $H_2$ asymptote. The wave packet was propagated using the modified Chebyshev recursion relation:[55–57]

$$T_{k+1}(\hat{H}_s) = 2D\hat{H}_s\, T_k(\hat{H}_s) - D^2 T_{k-1}(\hat{H}_s), \qquad k \geq 1 \tag{S28}$$

with $T_0(\hat{H}_s) = 1$ and $T_1(\hat{H}_s) = \hat{H}_s$. $D$ is the damping function (given in Table S2), which is used to avoid reflections at the edge of the grid. In addition, the Hamiltonian needs to be scaled to the spectral range $(-1, 1)$ via

$$\hat{H}_s = \frac{\mathbf{H} - H^+ \mathbf{I}}{H^-}, \tag{S29}$$

in which the spectral midpoint $H^+ = (H_{\max} + H_{\min})/2$ and the half-width $H^- = (H_{\max} - H_{\min})/2$ were determined by the spectral extrema, $H_{\max}$ and $H_{\min}$, which can be readily estimated. As the wave packet is propagated to the product asymptote, the energy dependent $S$-matrix can then be accumulatively computed by:[60,62]

$$S^{J_p}_{f \leftarrow i}(E) = \frac{1}{2\pi H^- \sin\vartheta a_i(E) a_f^*(E)} \left\langle \psi_f^- \left| \sum_{k=0}^{+\infty} (2 - \delta_{k0}) e^{-ik\vartheta} T_k(\hat{H}_s) \right| \psi_i^+ \right\rangle \tag{S30}$$

where $E$ is the total energy, $\vartheta = \arccos(E)$ is the Chebyshev angle, and $p$ denotes the parity. $a_i$ and $a_f$ are the energy amplitudes of the initial reactant and final product wave packets $\psi_i$ and $\psi_f$, respectively.

Once the $S$-matrix ( $S^J_{v_f j_f K_f \leftarrow v_i j_i K_i}(E)$) elements are obtained, the state-to-state differential cross section (DCS) can be computed as[63]

$$\frac{d\sigma_{v_f j_f \leftarrow v_i j_i}(\theta, E)}{d\Omega} = \frac{\pi}{2j_i + 1} \sum_{K_i} \sum_{Kf} \left| \frac{1}{2ik_{v_i j_i}} \sum_J (2J+1) d^J_{K_i K_f}(\pi - \theta) S^J_{v_f j_f K_f \leftarrow v_i j_i K_i}(E) \right|^2 \tag{S31}$$

where $\theta$ is the scattering angle in the space-fixed frame and $d^J_{K_f K_i}$ is the reduced Wigner rotational matrix. Extensive convergence tests were carried out to determine the optimal numerical parameters (shown in Table S2) for the quantum dynamics calculations. In particular, a maximum $K$ (the projection of the total angular momentum $J$on the $R$ axis) value of 14 was sufficient for all partial waves for $Ar^+(^2P_{3/2})$ collisions with $H_2(v'' = 0,\ j'' = 0/1)$.

## Trajectory surface hopping dynamics

The mixed quantum-classical trajectory surface hopping (TSH) simulations was performed using the fewest switches with time uncertainty (FSTU) method,[64] as implemented in the ANT program.[65] Nonadiabatic dynamics were propagated in the diabatic representation using a stochastic decoherence scheme[66] combined with FSTU. All frustrated hops were treated according to the grad$V$ prescription.[67] The initial state of $H_2$ was set to the rovibrational ground state ($v'' = 0, j'' = 0$) in channel 1, with an initial intermolecular separation of 8.0 Å between the collision partners. Trajectories were terminated upon reaching a separation of 8.5 Å. The impact parameter $b$ was sampled via the relation $b = b_{\max}\zeta^{1/2}$, where $\zeta$ is a uniformly distributed random variable in [0, 1] and $b_{\max}$ was set to the initial reactant separation of 8.0 Å. The rotational quantum number was obtained by rounding the rotational angular momentum to the nearest integer.

## Dynamics results at the collision energy of 0.22 eV

Figure S4(b) presents the sliced $H_2^+$ product scattering image for the charge transfer reaction $Ar^+(^2P_{3/2}) + H_2 \rightarrow Ar + H_2^+(v', j')$ at the collision energy of 0.22 eV measured with the Beijing setup. The three red concentric rings with numbers represent the kinematic cutoffs for $H_2^+(v' = 0, 1$ and $2)$, respectively.

The corresponding product internal energy distributions are shown in Figures S4(c) (black solid curve for the scattering angle range of 0–10° in the forward scattering direction) and S4 (d) (black solid curve for the scattering angle range of 0–180°). Similar to the experimental results at the collision energy of 0.04 eV (shown in Figure 1), the most interesting observation here is that the $H_2^+(v' = 0, 1)$ product is highly rotationally excited and shows prominent non-statistical characters, despite the lack of definitive rotational state resolution.

Moreover, these highly rotationally excited products are strongly forward peaked with a minor weak peak in the backward region, as shown by the sliced product image in Figure S4(b). As shown in Figures S4(c) and S4(d), the calculated product internal energy distributions (considering both initial $H_2(v'' = 0, j'' = 0, 1)$ states) are in satisfactory agreement with the experimental measurements in the forward scattering region of 0–10° and the full scattering angle range of 0–180°.

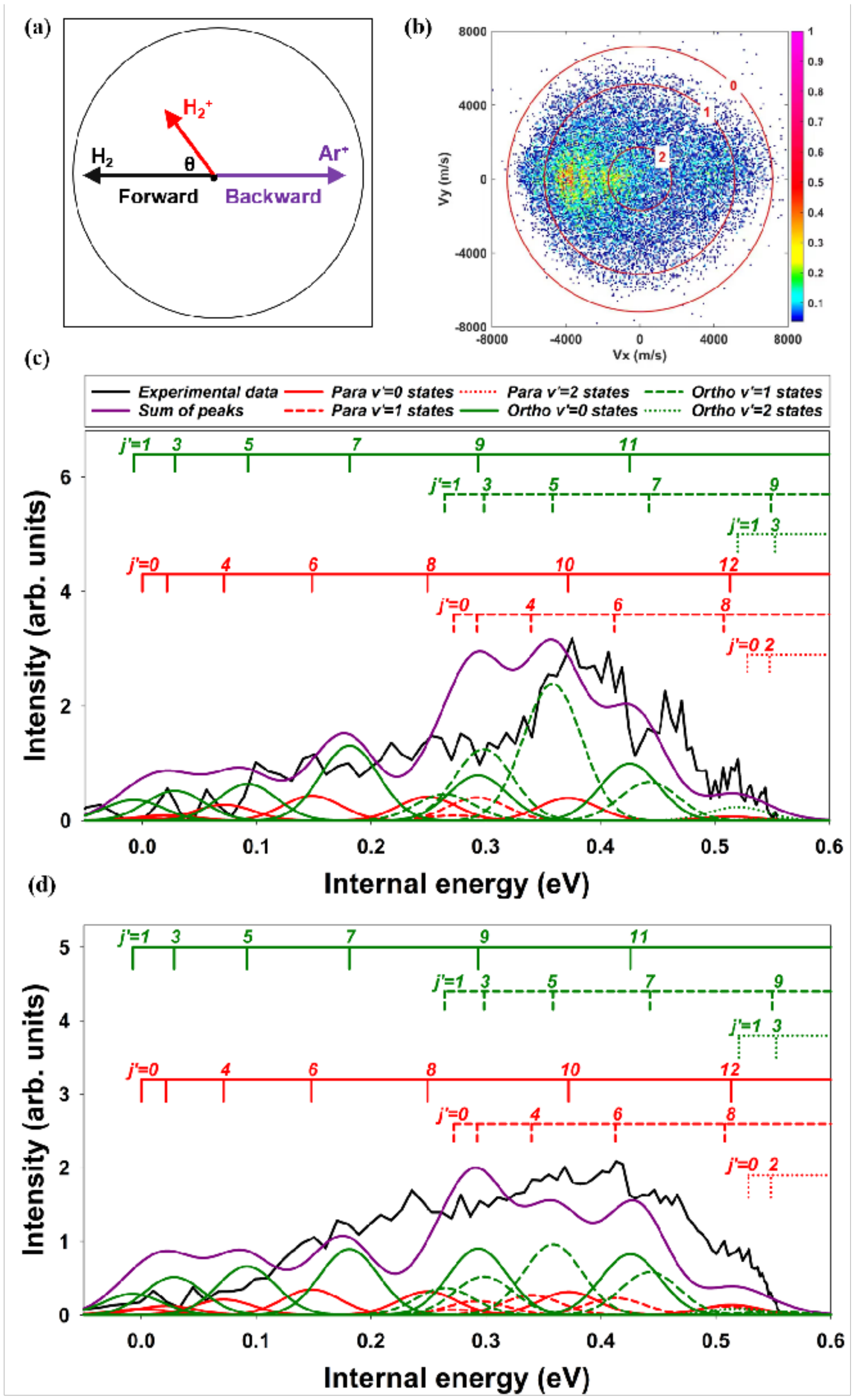

Figure S4: $Ar^+(^2P_{3/2})$ + $H_2$ → Ar + $H_2^+(v', j')$ at the collision energy of 0.22 eV: (a) Newton diagram of the reaction; (b) Sliced $H_2^+$ product scattering image, the red rings with numbers represent the kinematic cutoffs for $H_2^+(v'$=0, 1 and 2), respectively, the red arrows indicate the directions of $H_2$ and $Ar^+$ in the COM frame; (c) $H_2^+$ product internal energy distributions in the forward scattering region (scattering angle range of 0–10°) (black: experiment, red, green, and purple: quantum calculations (QM) for $j''$=0,1 and the statistical average 25%:75%, markers denote the energies of the product quantum states); (d) the full $H_2^+$ product internal energy distributions (scattering angle range of 0–180°). The QM distributions were generated by Gaussian convolution with a fixed FWHM of 0.06 eV to well match the experimental distributions.

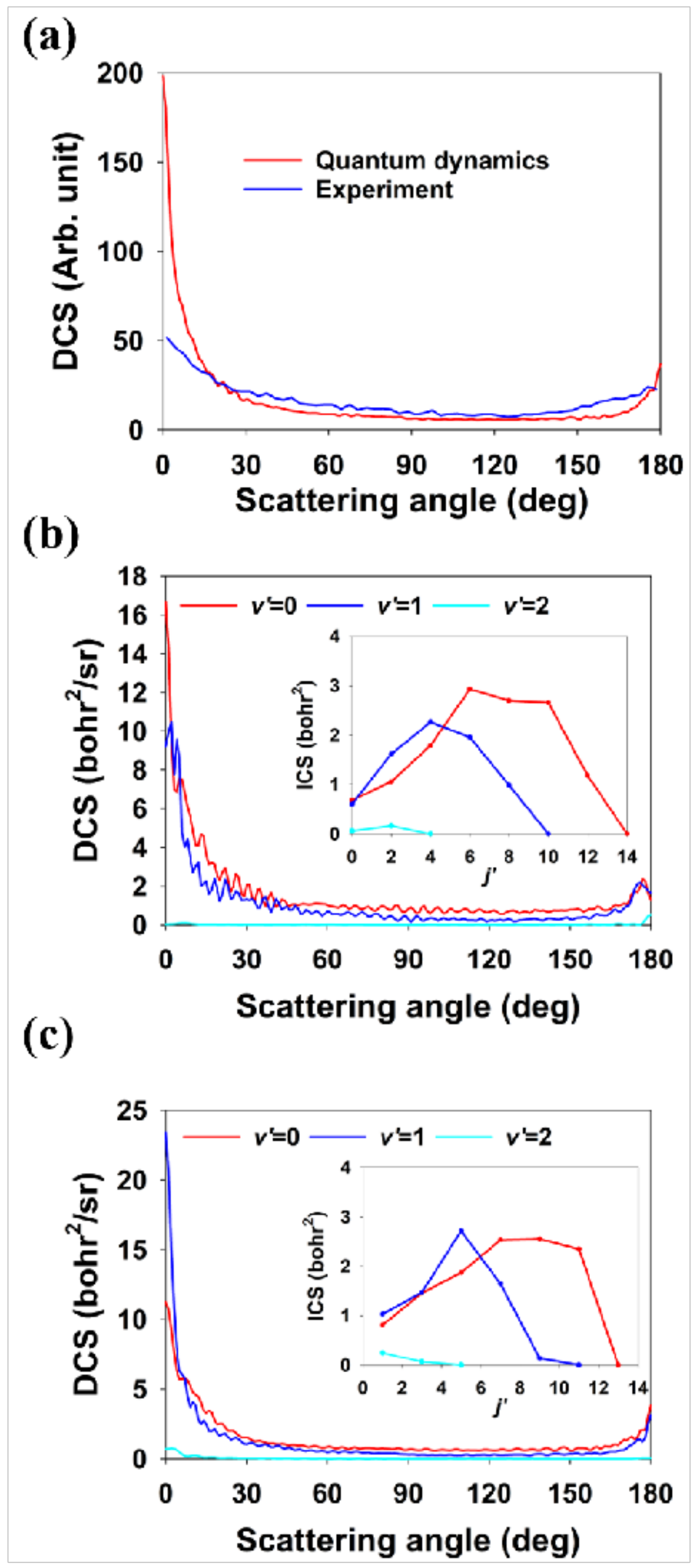

Figure S5: Quantum scattering results for the charge transfer reaction $Ar^+(^2P_{3/2})$ + $H_2(v'' = 0, j'' = 0, 1) \rightarrow$ Ar + $H_2{}^+(v', j')$ at the collision energy of 0.22 eV. (a) Total DCSs (red line: QM calculations; blue: experiment); product vibrational state-resolved DCSs with product rotational distributions shown in the inset for $j'' = 0$ (b) and $j'' = 1$ (c).

Figure S5 displays the calculated DCSs by the QM method at the collision energy of 0.22 eV for different product vibrational states. The $H_2{}^+$ products are strongly forward peaked for all the internal energies of $H_2{}^+$, and a minor weak peak is also obvious in the backward direction, which is in satisfactory agreement with the experimental observations shown in Figure S5(a). The ${H_2}^+(v' = 0)$ and ${H_2}^+(v' = 1)$ rotational levels are found to be peaked at $j' \sim 8$ and $\sim 4$, respectively, for the initial $H_2(v'' = 0, j'' = 0)$ state and $j' \sim 9$ and $\sim 5$, respectively, for the initial $H_2(v'' = 0, j'' = 1)$ state, showing prominent non-statistical characters, which is also consistent with the experimental observation.

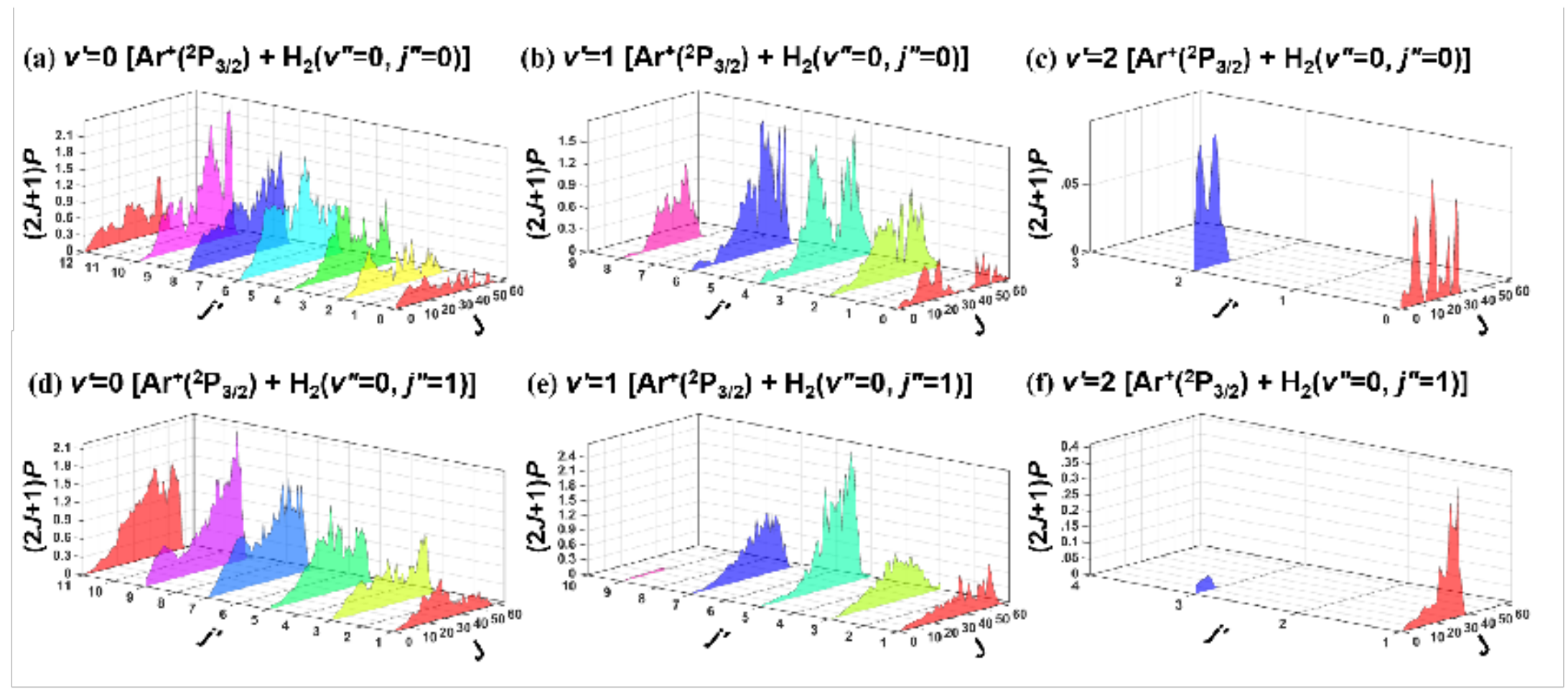

Figure S6: Opacity functions at the collision energy of 0.22 eV for $H_2(v'' = 0, j'' = 0, 1)$ excited to various product vibrational and rotational levels.

Figure S6 shows quantum opacity functions for excitations from para-/ortho-$H_2$ $(v'' = 0, j'' = 0, 1)$ via collisions with $Ar^+(^2P_{3/2})$ to various final ${H_2}^+$ vibrational and rotational states. Consistent with the results at the collision energy of 0.04 eV shown in Figure 3, the ${H_2}^+$ products with higher rotational excitation have larger contributions from higher angular momentum $J$ partial waves. Furthermore, all the opacity functions, especially for the highly rotationally excited products, show abrupt cutoffs at $J \sim 60$, consistent with the character predicted by the capture model.

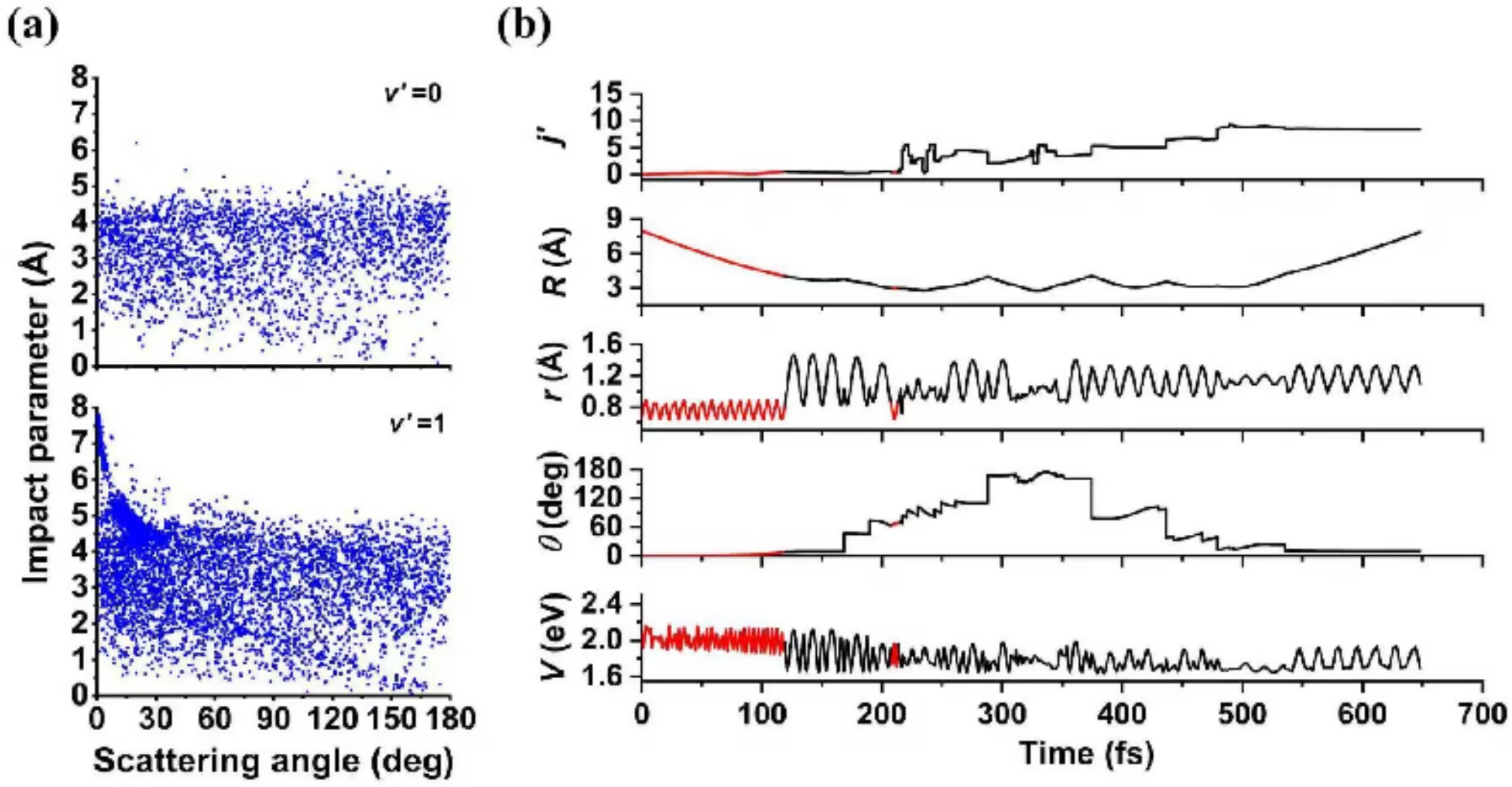

Figure S7: Trajectory surface-hopping results at the collision energy of 0.22 eV. (a) Correlation between the impact parameter and scattering angle; (b) evolution of various attributes as a function of time for an exemplary trajectory: rotational level $j'$, Jacobi radius $R$, H-H bond length $r$, scattering angle $\theta$, and potential energy V (red: $Ar^+$ ($^2P_{3/2}$ ) + $H_2$; black: Ar + $H_2^+$ ) .

Additional insights come from the TSH calculations performed at the collision energy of 0.22 eV. Figure S7(a) shows the correlation between the scattering angle and the impact parameter. Similar to the low collision energy case shown in Figure 4, the lack of correlation is suggestive of the formation of a complex. In the same figure, an exemplary trajectory with an impact parameter of 4 Å is shown to scatter in the forward direction ($\theta = 9°$) with rotational excitation to $j' = 8$. It is clear that the charge transfer occurs at $\sim$ 120 fs, and then the two colliders are trapped in the ground-state potential well for a significant time ($\sim$ 400 fs), orbiting with each other, very similar to the low collision energy case described in the main text. This orbiting process renders the translational energy gradually converted to the rotational excitation of the product, as shown in the top panel of Figure S7(b). The trajectory also shows that the system does not reach the short-range repulsive wall on the PES. These trajectories revealed that the collisional rotational excitation at both collision energies (0.04 and 0.22 eV) is facilitated mostly by the anisotropy of the attractive potential in the ground electronic state, as shown in Figure 2(a).

# Tables

## Table S1: Energy-dependent weights in NN fitting of DPEM

| Energy range (eV) | Weight |
|---|---|
| $< 4$ | 1.0 |
| 4–5 | 0.5 |
| 5–6 | 0.2 |
| 6–7 | 0.1 |
| 7–9 | 0.03 |
| 9–13 | 0.01 |

## Table S2: Numerical parameters used in quantum scattering calculations for the charge transfer reaction $\mathbf{Ar^+(^2P_{3/2}) + H_2}(v''=0, j''=0,1) \rightarrow \mathbf{Ar + H_2{}^+}(v', j')$. Atomic units used unless stated otherwise.

$R \in [2, 21.0], N_R = 175,$

$r \in [0.6, 7.0], N_r = 55,$

$\gamma \in (0, 180), j = 0 \sim 49,\ N_j = 50.\ K_{\max} = 14,\ J_{\max} = 65.$

Initial wave packet:

$\exp[-(R-R_0)^2/2\Delta^2]cos(k_0R), k_0 = \sqrt{2E_0\mu_R}$

$R_0 = 16.0,\ E_0 = 0.3$ eV, $\Delta = 0.3;$

Damping function $D$:

$\exp[-0.008(R-19.5)^2]$,for $19.5 < R < 21.0$

$\exp[-0.005(R-16.5)^2]$,for $16.5 < R < 19.5,$

$\exp[-0.25(r-4.5)^2]$,for $4.5 < r < 7.0;$

$= 1$ otherwise.

Analysis plane:

$R = 15.0$.

Chebyshev steps:

20000.